# Full-Rate, Full-Diversity, Finite Feedback Space-Time Schemes with Minimum Feedback and Transmission Duration

Lakshmi Prasad Natarajan and B. Sundar Rajan, *Senior Member, IEEE*

*Abstract*—In this paper a MIMO quasi static block fading channel is considered in which the transmitter has partial channel knowledge obtained via a finite $N$-ary delay-free, noise-free feedback from the receiver. The transmitter uses a set of $N$ Space-Time Block Codes (STBCs), one corresponding to each of the $N$ possible feedback values, to encode and transmit information bits. The particular feedback function used at the receiver and the $N$ component STBCs used at the transmitter together constitute a *Finite Feedback Scheme (FFS)*. If each of the component codes encodes $K$ independent complex symbols and is of transmission duration $T$, the rate of the FFS is $\frac{K}{T}$ complex symbols per channel use. Although a number of FFSs are available in the literature that provably achieve full-diversity, such as transmit antenna selection, beamforming, and precoding of STBCs, there is no known universal criterion to determine whether a given arbitrary FFS achieves full-diversity or not. Further, all known full-diversity FFSs for $T < N_t$ where $N_t$ is the number of transmit antennas, have rate at the most 1. In this paper a universal necessary condition for any FFS to achieve full-diversity is given, using which the notion of *Feedback-Transmission duration optimal (FT-Optimal)* FFSs - schemes that use minimum amount of feedback $N$ given the transmission duration $T$, and minimum transmission duration given the amount of feedback to achieve full-diversity - is introduced. When there is no feedback, i.e., when $N = 1$, an FT-optimal scheme consists of a single STBC with $T = N_t$, and the universal necessary condition reduces to the well known necessary and sufficient condition for an STBC to achieve full-diversity viz., every non-zero codeword difference matrix of the STBC must be of rank $N_t$. Also, a sufficient condition for full-diversity is given for the class of FFSs in which the feedback chooses the component STBC with the largest minimum Euclidean distance. Using this sufficient condition full-rate (rate $N_t$) full-diversity FT-Optimal schemes are constructed for all triples $(N_t, T, N)$ with $NT = N_t$. These are the first full-rate full-diversity FFSs reported in the literature for $T < N_t$. Finally, simulation results are presented that show that the new FFSs have the best error performance among all the schemes available in the literature.

## I. INTRODUCTION

We consider quasi-static block fading multiple-input multiple-output (MIMO) wireless channel with Rayleigh flat fading. We assume that the receiver has full-channel state information, and the transmitter has only a partial knowledge of the channel obtained through a delay-free noise-free $N$-ary feedback index conveyed by the receiver. The transmitter is equipped with $N$ Space-Time Block Codes (STBCs), one corresponding to each of the $N$ different values of the feedback index, and based on the received feedback value, it uses the corresponding STBC to encode and transmit information bits. The receiver, knowing the feedback index that it has sent to the transmitter and hence the STBC used for encoding, performs maximum-likelihood (ML) decoding of transmitted codeword to estimate the information bits. The feedback function used by the receiver to generate the $N$-ary feedback index, and the $N$ component STBCs used by the transmitter determine the communication protocol implemented on the MIMO channel with feedback. Throughout this paper we will refer to the combination of the particular feedback function used at the receiver with the $N$ component STBCs used at the transmitter as a *Finite Feedback Scheme (FFS)*. If each of the component STBCs encodes $K$ independent complex symbols and has transmission duration $T$, we say that the FFS has rate $R = \frac{K}{T}$ complex symbols per channel use. The definition of FFS is universal and subsumes all schemes available in the literature for delay-free noise-free finite feedback channels with quasi-static block fading, such as transmit antenna selection [1], precoding for spatial multiplexing systems [2], beamforming [3]–[6], combining space-time codes with beamforming [7]–[9], extending orthogonal STBCs [10], switching between orthogonal STBC and spatial multiplexing [11], and code diversity [12] (See Section II-A for formal definition of an FFS, and Table I for a summary of some of the FFSs available in the literature).

A number of FFSs are available in the literature that provably achieve full-diversity such as transmit antenna selection [1] and the schemes in [4]–[12]. However, there is no known universal criterion (applicable to any finite feedback scheme, including those in [1]–[12] as special cases) to determine whether a given arbitrary FFS achieves full-diversity or not. Further, all known full-diversity FFSs for $T < N_t$, where $N_t$ is the number of transmit antennas, have rate at the most 1. In this context the contributions (and organization) of this paper are as follows.

- We first give a universal necessary condition for any FFS to achieve full-diversity (Corollary 1, Section II-B). Using this necessary condition we introduce the notion of *Feedback-Transmission duration optimal (FT-Optimal)* FFSs - schemes that use minimum amount of feedback given the transmission duration and minimum transmission duration given the amount of feedback to achieve

The authors are with the Department of Electrical Communication Engineering, Indian Institute of Science, Bangalore-560012, India (e-mail: {nlp, bsrajan}@ece.iisc.ernet.in).

full-diversity. The class of FT-optimal FFSs consists of all full-diversity schemes for which the product of feedback index set cardinality $N$ and transmission duration $T$ equals the number of transmit antennas $N_t$. When there is no feedback, i.e., when $N = 1$, an FT-optimal scheme consists of a single STBC with $T = N_t$, and the universal necessary condition reduces to the well known necessary and sufficient condition for an STBC to achieve full-diversity viz., every non-zero codeword difference matrix of the STBC must be of rank $N_t$ (Section II-B).

- For FFSs which use the feedback function that chooses the component STBC with the largest minimum Euclidean distance, we give a sufficient condition for full-diversity (Theorem 2, Section II-C).
- Using the sufficient criterion and tools from algebraic number theory we construct full-rate (rate $R = N_t$) full-diversity FT-Optimal schemes for all triples $(N_t, T, N)$ with $NT = N_t$ (Section III). These are the first full-rate full-diversity FFSs reported in the literature for $T < N_t$.
- We present simulation results comparing the bit error rate performance of the new schemes with the schemes already available in the literature which show that the new FFSs have the best performance while utilizing the least amount of feedback and transmission duration (Section IV).

The system model is explained in Section II-A, the definitions and results from algebraic number theory that we have used in this paper are briefly reviewed in Section III-A, and finally the paper is concluded in Section V.

*Notation:* Throughout the paper, matrices (column vectors) are denoted by bold, uppercase (lowercase) letters. For a complex matrix $\mathbf{A}$, the transpose, the conjugate-transpose and the Frobenius norm are denoted by $\mathbf{A}^T$, $\mathbf{A}^H$ and $||\mathbf{A}||_F$ respectively. For a square matrix $\mathbf{A}$, $\det(\mathbf{A})$ is the determinant of $\mathbf{A}$, and $\text{tr}(\mathbf{A})$ is the trace of $\mathbf{A}$. For any positive integer $n$, $\mathbf{I}_n$ is the $n \times n$ identity matrix, and $\mathbf{0}$ is the all zero matrix of appropriate dimension. Unless used as a subscript $i$ denotes $\sqrt{-1}$. The indicator function is denoted by $\mathbf{1}(\cdot)$, and for any vector $\mathbf{u}$, its $\ell^{th}$ component is denoted by $\mathbf{u}(\ell)$.

## II. FULL-DIVERSITY CRITERIA: A UNIVERSAL NECESSARY CONDITION, AND A SUFFICIENT CONDITION

### A. System Model

We consider an $N_t \times N_r$ quasi-static Rayleigh flat fading MIMO channel $\mathbf{Y} = \sqrt{E}\mathbf{X}\mathbf{H} + \mathbf{W}$, where $\mathbf{Y}$ is the $T \times N_r$ received matrix, $\mathbf{X}$ is the $T \times N_t$ transmit matrix, $\mathbf{H}$ is the $N_t \times N_r$ channel matrix, $\mathbf{W}$ is the $T \times N_r$ matrix representing the additive noise at the receiver and $E$ is the average transmit power. The entries of $\mathbf{H}$ and $\mathbf{W}$ are independent, zero mean, circularly symmetric complex Gaussian random variables, with the variance of each entry of $\mathbf{H}$ being 1, and the variance of each entry of $\mathbf{W}$ being $N_0$. The receiver uses a feedback function $\mathsf{f} : \mathbb{C}^{N_t \times N_r} \to \{1, \ldots, N\}$ to send the feedback index $\mathsf{f}(\mathbf{H})$ to the transmitter through a delay-free, noise-free feedback channel. A *Space-Time Block Code (STBC)* $\mathcal{C}$ is a finite set of $T \times N_t$ complex matrices. The transmitter is equipped with $N$ STBCs $\mathcal{C}_1, \ldots, \mathcal{C}_N$, with $|\mathcal{C}_1| = \cdots = |\mathcal{C}_N|$, one corresponding to each of the $N$ possible feedback indices. When $\mathsf{f}(\mathbf{H}) = n$, the transmitter uses the code $\mathcal{C}_n$ to encode the information bits. Upon receiving $\mathbf{Y}$, knowing the feedback index, and hence knowing the codebook used for transmission, the receiver performs ML decoding

$$\hat{\mathbf{X}} = \arg \min_{\mathbf{X} \in \mathcal{C}_{\mathsf{f}(\mathbf{H})}} ||\mathbf{Y} - \sqrt{E}\mathbf{X}\mathbf{H}||_F^2. \quad (1)$$

*Definition 1:* A *Finite Feedback Scheme (FFS)* for an $N_t \times N_r$ MIMO channel with $N$-ary noise-free, delay-free feedback and transmission duration $T$ is a tuple $(\mathsf{f}, \mathcal{C}_1, \ldots, \mathcal{C}_N)$, where $\mathsf{f} : \mathbb{C}^{N_t \times N_r} \to \{1, \ldots, N\}$ is the feedback function, and $\mathcal{C}_1, \ldots, \mathcal{C}_N$ are the $T \times N_t$ STBCs corresponding to each of the $N$ feedback indices.

*Example 1:* The FFS of [6], known as Grassmannian beamforming, is of transmission duration $T = 1$. The transmitter is equipped with $N$ unit norm vectors $\mathbf{u}_1, \ldots, \mathbf{u}_N \in \mathbb{C}^{N_t \times 1}$ known as the *beamforming vectors*. Let $\mathcal{A} \subset \mathbb{C}$ be a finite signal set such as QAM, HEX or a PSK constellation. Then, for $n \in \{1, \ldots, N\}$, the $n^{th}$ component STBC of the FFS from [6] is $\mathcal{C}_n = \{a\mathbf{u}_n^T | a \in \mathcal{A}\}$. The feedback function used is $\mathsf{f}(\mathbf{H}) = \arg\max_{n \in \{1, \ldots, N\}} ||\mathbf{u}_n^T \mathbf{H}||_F^2$. ∎

Table I summarizes some of the FFSs available in the literature. The scheme from [11] uses two codes of different rates: the Alamouti code [14] with rate 1 and spatial multiplexing with rate 2, hence the rate of this FFS is not defined. The last row corresponds to $N = 1$, i.e., MIMO channels without feedback. In this case the FFS consists of a single STBC $\mathcal{C}_1$, and the feedback value is equal to 1 for all $\mathbf{H} \in \mathbb{C}^{N_t \times N_r}$.

An FFS is said to achieve a *diversity order* $d$ if the probability of decoding error $\mathsf{P}_e$ at the receiver decays as $\left(\frac{E}{N_0}\right)^{-d}$ i.e., if there exists a constant $c > 0$ such that $\mathsf{P}_e \leq c \left(\frac{E}{N_0}\right)^{-d}$, and an FFS is of *full-diversity* if it achieves a diversity order of $N_t N_r$.

If an STBC encodes $K$ independent complex symbols, its rate is $\frac{K}{T}$ complex symbols per channel use. The FFS $(\mathsf{f}, \mathcal{C}_1, \ldots, \mathcal{C}_N)$ is said to be of rate $R$ if each of the $N$ STBCs $\mathcal{C}_1, \ldots, \mathcal{C}_N$ is of rate $R$, and the FFS is of *full-rate* if $R = N_t$.

### B. A Universal Necessary Condition

Some notations are introduced before stating the criterion. For any STBC $\mathcal{C}$, let $\Delta\mathcal{C}$ denote the set of non-zero codeword difference matrices, i.e.,

$$\Delta\mathcal{C} = \{\mathbf{X}_1 - \mathbf{X}_2 \mid \mathbf{X}_1, \mathbf{X}_2 \in \mathcal{C}, \mathbf{X}_1 \neq \mathbf{X}_2\}.$$

For a given FFS $\mathcal{S} = (\mathsf{f}, \mathcal{C}_1, \ldots, \mathcal{C}_N)$ define the set $\Delta\mathcal{S}$ of $NT \times N_t$ matrices as

$$\Delta\mathcal{S} = \left\{ \begin{bmatrix} \mathbf{X}_1 \\ \mathbf{X}_2 \\ \vdots \\ \mathbf{X}_N \end{bmatrix} \middle| \mathbf{X}_1 \in \Delta\mathcal{C}_1, \ldots, \mathbf{X}_N \in \Delta\mathcal{C}_N \right\},$$

i.e., $\Delta\mathcal{S}$ is the set of all combinations of $N$ non-zero codeword difference matrices, one corresponding to each of the $N$ codes,

3TABLE I
EXAMPLES OF FINITE FEEDBACK SCHEMES AVAILABLE IN THE LITERATURE
(See table footnotes for notation.)

| Scheme | Setting | Component Code $\mathcal{C}_n$ | Feedback function $f(\mathbf{H})$ | Rate $R$ |
|---|---|---|---|---|
| Antenna Selection [1] | $N_t > 1$, $N = N_t$, $T = 1$, $\mathbf{e}_1, \ldots, \mathbf{e}_{N_t}$ are columns of $\mathbf{I}_{N_t}$ | $\{s\mathbf{e}_n^T \mid s \in \mathcal{A}\}$ | $\arg\max_{n \in \{1,\ldots,N\}}$ $\|\mathbf{e}_n^T \mathbf{H}\|_F^2$ | 1 |
| Precoded Spatial-Multiplexing [2] | $N_t, N > 1$, $T = 1$, $M < N_t$, $\mathbf{F}_1, \ldots, \mathbf{F}_N \in \mathbb{C}^{M \times N_t}$ | $\{\mathbf{s}^T \mathbf{F}_n \mid \mathbf{s} \in \mathcal{A}^M\}$ | $f_d(\mathbf{H})$ / $\arg\max_{n \in \{1,\ldots,N\}} \lambda_{\min}(\mathbf{F}_n \mathbf{H})$ / $\arg\max_{n \in \{1,\ldots,N\}} \det(\mathbf{I}_M + \frac{E}{N_0} \mathbf{F}_n \mathbf{H} \mathbf{H}^H \mathbf{F}_n^H)$ | $M$ |
| Heath, Jr. & Paulraj [4] | $N_t = 2$, $N > 1$, $T = 1$, $\gamma_n = e^{i\frac{2\pi n}{N}}$, $n \in \{1, \ldots, N\}$ | $\{[s \ s\gamma_n] \mid s \in \mathcal{A}\}$ | $\arg\max_{n \in \{1,\ldots,N\}}$ $\| [1 \ \gamma_n]^T \mathbf{H} \|_F^2$ | 1 |
| Grassmannian Beamforming [6] | $N_t, N > 1$, $T = 1$, $\mathbf{u}_1, \ldots, \mathbf{u}_N \in \mathbb{C}^{N_t \times 1}$ have unit norm | $\{s\mathbf{u}_n^T \mid s \in \mathcal{A}\}$ | $\arg\max_{n \in \{1,\ldots,N\}}$ $\|\mathbf{u}_n^T \mathbf{H}\|_F^2$ | 1 |
| Precoded Orthogonal STBCs [8] | $N_t, N > 1$, $M < N_t$, $\mathcal{C}$ is a $T \times M$ rate $R$ orthogonal STBC, $\mathbf{F}_1, \ldots, \mathbf{F}_N \in \mathbb{C}^{M \times N_t}$ | $\{\mathbf{X}\mathbf{F}_n \mid \mathbf{X} \in \mathcal{C}\}$ | $\arg\max_{n \in \{1,\ldots,N\}}$ $\|\mathbf{F}_n \mathbf{H}\|_F^2$ | $\leq 1$ |
| Heath, Jr. & Paulraj [11] | $N_t = N = T = 2$, $\|\mathcal{A}\| = \|\mathcal{A}'\|^2$ | $\mathcal{C}_1$ is Alamouti code using $\mathcal{A}$ $\mathcal{C}_2 = \left\{ \begin{bmatrix} s_1 & s_2 \\ s_3 & s_4 \end{bmatrix} \middle\| s_i \in \mathcal{A}' \right\}$ | $f_d(\mathbf{H})$ | NA |
| No feedback [13] | $N = 1$, $N_t, T \geq 1$ | $\mathcal{C}_1 \subset \mathbb{C}^{T \times N_t}$ | 1 | $\leq N_t$ |

Notation:
- $\mathcal{A}, \mathcal{A}' \subset \mathbb{C}$ are complex constellations such as QAM, HEX or PSK.
- $f_d(\mathbf{H}) = \arg\max_{n \in \{1,\ldots,N\}} \left\{ \min_{\mathbf{X} \in \Delta \mathcal{C}_n} \|\mathbf{X}\mathbf{H}\|_F^2 \right\}$, where $\Delta \mathcal{C}_n = \{\mathbf{X}_1 - \mathbf{X}_2 \mid \mathbf{X}_1, \mathbf{X}_2 \in \mathcal{C}_n, \mathbf{X}_1 \neq \mathbf{X}_2\}$.
- $\lambda_{\min}(\mathbf{A})$ is the smallest singular value of $\mathbf{A}$.

stacked on top of one another. Further, let

$$r(\Delta \mathcal{S}) = \min\{\text{rank}(\mathbf{X}) \mid \mathbf{X} \in \Delta \mathcal{S}\}.$$

Since the matrices in the set $\Delta \mathcal{S}$ are of dimension $NT \times N_t$, we have $r(\Delta \mathcal{S}) \leq N_t$.

*Theorem 1:* An FFS $\mathcal{S}$ achieves a diversity order of at the most $r(\Delta \mathcal{S}) N_r$.

*Proof:* Proof is given in Appendix A. ■

The following necessary condition for full-diversity follows immediately from the above theorem.

*Corollary 1:* If an FFS $\mathcal{S}$ achieves full-diversity, then $r(\Delta \mathcal{S}) = N_t$ and $NT \geq N_t$.

*Proof:* Since $\mathcal{S}$ achieves full-diversity, from Theorem 1, $N_t N_r \leq r(\Delta \mathcal{S}) N_r$ i.e., $r(\Delta \mathcal{S}) \geq N_t$. But $\Delta \mathcal{S}$ is a set of $NT \times N_t$ matrices, and the matrices belonging to $\Delta \mathcal{S}$ can have rank at the most equal to $N_t$, thus we have $r(\Delta \mathcal{S}) = N_t$. It follows that the rank of each $\mathbf{X} \in \Delta \mathcal{S}$ is $N_t$ and hence the number of rows of $\mathbf{X}$ $NT \geq N_t$. ■

*Example 2:* Continuing with Example 1, we have that $\Delta \mathcal{C}_n = \{a\mathbf{u}_n^T \mid a \in \Delta \mathcal{A}\}$, where $\Delta \mathcal{A} = \{a_1 - a_2 \mid a_1, a_2 \in \mathcal{A}, a_1 \neq a_2\}$. Each member of $\Delta \mathcal{S}$ is a matrix of the form $[a_1 \mathbf{u}_1 \ a_2 \mathbf{u}_2 \ \cdots \ a_N \mathbf{u}_N]^T$, where $a_1, a_2, \ldots, a_N \in \Delta \mathcal{A}$ and hence are non-zero. This matrix will have rank $N_t$ if and only if the linear span of the vectors $\mathbf{u}_1, \ldots, \mathbf{u}_N$ is $\mathbb{C}^{N_t \times 1}$. In [6] it is shown that this is also a sufficient condition for this scheme to attain full-diversity. ■

From Corollary 1, for a scheme to achieve full-diversity the product of its transmission duration and the cardinality of feedback index set must be at least $N_t$.

*Definition 2:* A full-diversity FFS is said to be *Feedback-Transmission duration optimal (FT-optimal)* if $NT = N_t$.

An FT-optimal scheme uses the minimum amount of feedback $N$ given the transmission duration $T$, and minimum transmission duration given the amount of feedback to attain full-diversity. When there is no feedback, i.e., when $N = 1$, an FT-optimal scheme consists of a single STBC with $T = N_t$, and the necessary condition of Corollary 1 reduces to the well known necessary and sufficient condition of [13] for an STBC to achieve full-diversity viz., every non-zero codeword difference matrix of the STBC must be of rank $N_t$. On the other hand, for the case of least possible transmission duration $T = 1$, an FT-optimal scheme uses an $N = N_t$-ary feedback. In Section III we construct FT-optimal schemes for all $N_t \geq 1$ and all pairs $(N, T)$ such that $NT = N_t$.

### C. A Sufficient Condition

Let $f_d(\mathbf{H})$ be the feedback function that returns the index of the codebook with largest minimum Euclidean distance for the given channel $\mathbf{H}$, i.e.,

$$\mathsf{f}_\mathsf{d}(\mathbf{H}) = \arg \max_{n \in \{1,\ldots,N\}} \left\{ \min_{\mathbf{X} \in \Delta \mathcal{C}_n} ||\mathbf{X}\mathbf{H}||_F^2 \right\}. \quad (2)$$

We now show that for any FFS that uses $\mathsf{f} = \mathsf{f}_\mathsf{d}$, the necessary condition of Corollary 1 is also a sufficient condition to achieve full-diversity.

*Theorem 2:* The FFS $\mathcal{S} = (\mathsf{f}_\mathsf{d}, \mathcal{C}_1, \ldots, \mathcal{C}_N)$ achieves full-diversity if $\mathsf{r}(\Delta \mathcal{S}) = N_t$.

*Proof:* See Appendix B. ∎

*1) A new full-diversity FFS:* As an example for the application of Theorem 2, we now construct a new $N = 2$, $T = 1$ FT-optimal, full-rate, full-diversity FFS for $N_t = 2$ antennas. Let $\mathbf{x}_1, \mathbf{x}_2$ be complex symbols encoded using a QAM constellation $\mathcal{A} \subset \mathbb{Z}[i]$. Let $\mathbb{Q}(i, \sqrt{5})$ be the field obtained from $\mathbb{Q}$ by the adjunction of elements $i = \sqrt{-1}$ and $\sqrt{5}$, and $\sigma : \mathbb{Q}(i, \sqrt{5}) \to \mathbb{Q}(i, \sqrt{5})$ be the automorphism on $\mathbb{Q}(i, \sqrt{5})$ that fixes $\mathbb{Q}(i)$ and maps $\sqrt{5}$ to $-\sqrt{5}$. Define

$$\mathcal{C}_1 = \left\{ \begin{bmatrix} \alpha(x_1 + x_2\theta) & \sigma(\alpha(x_1 + x_2\theta)) \end{bmatrix} \;\middle|\; x_1, x_2 \in \mathcal{A} \right\} \text{ and}$$

$$\mathcal{C}_2 = \left\{ \begin{bmatrix} \alpha(x_1 + x_2\theta) & i\sigma(\alpha(x_1 + x_2\theta)) \end{bmatrix} \;\middle|\; x_1, x_2 \in \mathcal{A} \right\},$$

where $\theta = \frac{1+\sqrt{5}}{2}$ and $\alpha = 1 + i - i\theta$.

The Golden code [15], which is a full-diversity STBC for 2 transmit antennas with large coding gain is $\mathcal{C}_{\mathsf{Golden}} =$

$$\left\{ \begin{bmatrix} \alpha(x_1 + x_2\theta) & i\sigma(\alpha(y_1 + y_2\theta)) \\ \alpha(y_1 + y_2\theta) & \sigma(\alpha(x_1 + x_2\theta)) \end{bmatrix} \;\middle|\; x_1, x_2, y_1, y_2 \in \mathcal{A} \right\}.$$

The codes $\mathcal{C}_1$ and $\mathcal{C}_2$ correspond to the two 'threads' of the Golden code - $\mathcal{C}_1$ is obtained from the entries on the main diagonal of $\mathcal{C}_{\mathsf{Golden}}$ and $\mathcal{C}_2$ from the entries in the off-diagonal.

*Lemma 1:* The FFS $\mathcal{S} = (\mathsf{f}_\mathsf{d}, \mathcal{C}_1, \mathcal{C}_2)$ achieves full-diversity.

*Proof:* We need to show that every $\mathbf{X} \in \Delta \mathcal{S}$ has full rank. Since both $\mathcal{C}_1$ and $\mathcal{C}_2$ are linear, for any given $\mathbf{X} \in \Delta \mathcal{S}$ there exist $[x_1 \; x_2]^T, [y_1 \; y_2]^T \in \mathbb{Z}[i]^2 \setminus \{\mathbf{0}\}$, such that

$$\mathbf{X} = \begin{bmatrix} \alpha(x_1 + x_2\theta) & \sigma(\alpha(x_1 + x_2\theta)) \\ \alpha(y_1 + y_2\theta) & i\sigma(\alpha(y_1 + y_2\theta)) \end{bmatrix}.$$

Since $x_1, x_2 \in \mathbb{Q}(i)$ and $\{1, \theta\}$ is a basis of $\mathbb{Q}(i, \sqrt{5})$ as a vector space over $\mathbb{Q}(i)$, we have that $x = \alpha(x_1 + x_2\theta) \neq 0$. Similarly, $y = \alpha(y_1 + y_2\theta) \neq 0$. Since $\det(\mathbf{X}) = ix\sigma(y) - y\sigma(x)$ and $\sigma^2$ is the identity map on $\mathbb{Q}(i, \sqrt{5})$, we have $\det(\mathbf{X}) = iz - \sigma(z)$, where $z = x\sigma(y) \in \mathbb{Q}(i, \sqrt{5}) \setminus \{0\}$. If $\mathbf{X}$ is not of full rank, $\det(\mathbf{X}) = 0$, i.e., $i = \frac{\sigma(z)}{z}$ for some $z \in \mathbb{Q}(i, \sqrt{5})$. This would imply that

$$i = \sigma(i) = \sigma\left(\frac{\sigma(z)}{z}\right) = \frac{z}{\sigma(z)} = \left(\frac{\sigma(z)}{z}\right)^{-1} = -i,$$

which is not true. Hence, $i \neq \frac{\sigma(z)}{z}$ for any $z \in \mathbb{Q}(i, \sqrt{5})$, and $\mathbf{X}$ is of full rank. ∎

## III. NEW FULL-RATE FULL-DIVERSITY FT-OPTIMAL FINITE FEEDBACK SCHEMES

In this section, using tools from algebraic number theory, we construct full-rate full-diversity FT-optimal FFSs with $\mathsf{f} = \mathsf{f}_\mathsf{d}$ for all parameters $N, T$ and $N_t$ such that $N_t = NT$. In Section III-A we briefly review some definitions and results from algebraic number theory which we use to construct new schemes in Section III-B ($T = 1$ case) and Section III-C ($T > 1$ case).

### A. Preliminaries

For any two fields $\mathbb{K}$ and $\mathbb{F}$, if $\mathbb{F} \subseteq \mathbb{K}$ then $\mathbb{K}$ is said to be an *extension* of $\mathbb{F}$, and $\mathbb{F}$ a *subfield* of $\mathbb{K}$. For any $\alpha \in \mathbb{K}$, $\mathbb{F}(\alpha)$ denotes the smallest subfield of $\mathbb{K}$ that contains $\mathbb{F}$ and $\alpha$, and it consists of all the elements of the form $\frac{f(\alpha)}{h(\alpha)}$, where $f, h \in \mathbb{F}[x]$ are polynomials over $\mathbb{F}$ and $h(x) \neq 0$. An element $\alpha \in \mathbb{C}$ is said to be an *algebraic number*, or simply *algebraic*, if there exists a non-zero polynomial $f \in \mathbb{Q}[x]$ such that $f(\alpha) = 0$. If $\alpha$ is algebraic, the field $\mathbb{Q}(\alpha)$ is said to be an *algebraic number field*.

*Example 3:* For any $a \in \mathbb{Q}$, $\sqrt{a}$ is algebraic, since it satisfies the equation $x^2 - a = 0$. Hence, $\sqrt{2}, \sqrt{3}, i = \sqrt{-1}$ are all algebraic. Also, $\frac{1+\sqrt{5}}{2}$ is algebraic since it is a root of the equation $x^2 - x - 1 = 0$. ∎

*Lemma 2 ([16, p. 107]):* The sum, difference, product and quotient of algebraic numbers are themselves algebraic numbers.

We will use the following result to prove the full-diversity property of our FFSs.

*Theorem 3 (Lindemann-Weierstrass Theorem [17, p. 6]):* If $\alpha_1, \ldots, \alpha_m$ are distinct algebraic numbers, and $c_1, \ldots, c_m$ are algebraic numbers that are not all equal to zero, then

$$c_1 e^{\alpha_1} + c_2 e^{\alpha_2} + \cdots + c_m e^{\alpha_m} \neq 0.$$

The following result gives a procedure to construct sets of algebraic numbers, of any desired finite cardinality, that are linearly independent over $\mathbb{Q}$. We will use this result to construct full-diversity FFSs for $T > 1$ in Section III-C.

*Theorem 4 ([18]):* Let $n_1, \ldots, n_m$ be positive integers, $p_1, \ldots, p_m$ be distinct primes, and $b_1, \ldots, b_m$ be positive integers not divisible by any of these primes. For $k = 1, \ldots, m$, let $\alpha_k = \sqrt[n_k]{b_k p_k}$, and $f(x_1, \ldots, x_m) \in \mathbb{Q}[x_1, \ldots, x_m]$ be any polynomial in indeterminates $x_1, \ldots, x_m$ with degree less than or equal to $n_k - 1$ with respect to $x_k$. Then, $f(\alpha_1, \ldots, \alpha_m) = 0$ if and only if all the coefficients of $f$ are equal to zero.

It follows immediately from the above theorem that the set

$$\left\{ \alpha_1^{\ell_1} \alpha_2^{\ell_2} \cdots \alpha_m^{\ell_m} \;\middle|\; 0 \leq \ell_k < n_k, \; k = 1, \ldots, m \right\},$$

with cardinality $\prod_{k=1}^{m} n_k$, is linearly independent over $\mathbb{Q}$. Note that the above set of algebraic numbers obtained from Theorem 4 is real. On multiplying each of the elements of this set with $i$, we get a set of purely imaginary algebraic numbers that are $\mathbb{Q}$-linearly independent. We are interested in purely imaginary numbers as these will lead to FFSs in Section III with the same average transmit energy per each transmit antenna.

*Example 4:* Let $m = 2$, $p_1 = 2$ and $p_2 = 3$ be the two distinct primes, and $b_1 = b_2 = 1$. Suppose we want a set of $n_1 n_2 = 4$ algebraic numbers that are linearly independent over

$\mathbb{Q}$. Choosing $n_1 = n_2 = 2$, we have $\alpha_1 = \sqrt{2}$ and $\alpha_2 = \sqrt{3}$. From Theorem 4,

$$\left\{ \alpha_1^{\ell_1} \alpha_2^{\ell_2} \mid 0 \leq \ell_1, \ell_2 < 2 \right\} = \left\{ 1, \sqrt{2}, \sqrt{3}, \sqrt{6} \right\}$$

is linearly independent over $\mathbb{Q}$. On multiplying each of the elements of the above set by $i$, we see that $\{i, i\sqrt{2}, i\sqrt{3}, i\sqrt{6}\}$ is linearly independent over $\mathbb{Q}$. ∎

In [19]–[22] rotation matrices $\mathbf{U} \in \mathbb{C}^{m \times m}$ where constructed for all $m > 1$ with non-zero minimum product distance, i.e., with the property that for any $\mathbf{a} \in \mathbb{Z}[i]^m \setminus \{\mathbf{0}\}$ and $\mathbf{s} = \mathbf{U}\mathbf{a}$, $\prod_{\ell=1}^{m} |\mathbf{s}(\ell)| > 0$, where $\mathbf{s}(\ell)$ denotes the $\ell^{th}$ component of $\mathbf{s}$. Further, these matrices were constructed over algebraic number fields, i.e., each component of $\mathbf{U}$ is an algebraic number. These matrices are known as *full-diversity algebraic rotations*, and a table of the best known (in terms of minimum product distance) full-diversity algebraic rotations is available in [23].

### B. New Finite Feedback Schemes with $T = 1$

Let $\mathbf{U} \in \mathbb{C}^{N_t \times N_t}$ be any full-diversity algebraic rotation, $\alpha \in \mathbb{C}$ be any non-zero algebraic number, and $\gamma = e^{\alpha}$. The proposed FT-optimal FFS uses $N = N_t$ component STBCs, $\mathcal{C}_1, \ldots, \mathcal{C}_{N_t} \subset \mathbb{C}^{1 \times N_t}$, each of which encodes $N_t$ independent QAM symbols as follows. Let $\mathbf{a} = [\mathbf{a}(1)\ \mathbf{a}(2)\ \cdots\ \mathbf{a}(N_t)]^T$ be a vector of $N_t$ independent symbols that take value from a QAM constellation $\mathcal{A} \subset \mathbb{Z}[i]$, and

$$\mathbf{s} = \begin{bmatrix} \mathbf{s}(1) & \mathbf{s}(2) & \cdots & \mathbf{s}(N_t) \end{bmatrix}^T = \mathbf{U}\mathbf{a}.$$

The $N_t$ component STBCs of the proposed FFS are

$$\mathcal{C}_1 = \left\{ \begin{bmatrix} \gamma \mathbf{s}(1) & \mathbf{s}(2) & \cdots & \mathbf{s}(N_t) \end{bmatrix} \mid \mathbf{s} = \mathbf{U}\mathbf{a},\ \mathbf{a} \in \mathcal{A}^{N_t} \right\},$$
$$\mathcal{C}_2 = \left\{ \begin{bmatrix} \mathbf{s}(1) & \gamma \mathbf{s}(2) & \cdots & \mathbf{s}(N_t) \end{bmatrix} \mid \mathbf{s} = \mathbf{U}\mathbf{a},\ \mathbf{a} \in \mathcal{A}^{N_t} \right\},$$
$$\vdots$$
$$\mathcal{C}_{N_t} = \left\{ \begin{bmatrix} \mathbf{s}(1) & \mathbf{s}(2) & \cdots & \gamma \mathbf{s}(N_t) \end{bmatrix} \mid \mathbf{s} = \mathbf{U}\mathbf{a},\ \mathbf{a} \in \mathcal{A}^{N_t} \right\}.$$
(3)

Each of the above STBCs is obtained from $\mathbf{s}^T$ by multiplying one of its components with $\gamma$. Note that the rate of the proposed scheme is $R = N_t$. Although the full-diversity property to be proved in Lemma 3 is valid for any non-zero algebraic $\alpha$, choosing $\alpha$ to be purely imaginary would ensure that $|\gamma| = 1$, and that for each of the component codes the average energy transmitted on each of the $N_t$ antennas is same.

*Example 5:* Consider the case $N_t = 3$. Using $\alpha = i\left(\frac{1+\sqrt{5}}{2}\right)$, $\gamma = e^{\alpha}$ and the $3 \times 3$ full-diversity rotation matrix

$$\mathbf{U} = \begin{bmatrix} -0.328 & -0.591 & -0.737 \\ -0.737 & -0.328 & 0.591 \\ -0.591 & 0.737 & -0.328 \end{bmatrix}$$

from [23], we get the following STBCs

$$\mathcal{C}_1 = \left\{ \begin{bmatrix} \gamma \mathbf{s}(1) & \mathbf{s}(2) & \mathbf{s}(3) \end{bmatrix} \mid \mathbf{s} = \mathbf{U}\mathbf{a},\ \mathbf{a} \in \mathcal{A}^{N_t} \right\},$$
$$\mathcal{C}_2 = \left\{ \begin{bmatrix} \mathbf{s}(1) & \gamma \mathbf{s}(2) & \mathbf{s}(3) \end{bmatrix} \mid \mathbf{s} = \mathbf{U}\mathbf{a},\ \mathbf{a} \in \mathcal{A}^{N_t} \right\} \text{ and}$$
$$\mathcal{C}_3 = \left\{ \begin{bmatrix} \mathbf{s}(1) & \mathbf{s}(2) & \gamma \mathbf{s}(3) \end{bmatrix} \mid \mathbf{s} = \mathbf{U}\mathbf{a},\ \mathbf{a} \in \mathcal{A}^{N_t} \right\}.$$
∎

*Lemma 3:* If $\mathbf{U}$ is a full-diversity algebraic rotation and $\alpha$ is a non-zero algebraic number, the FFS $\mathcal{S} = (\mathsf{f}_{\mathsf{d}}, \mathcal{C}_1, \ldots, \mathcal{C}_{N_t})$ achieves full-diversity, where $\mathcal{C}_1, \ldots, \mathcal{C}_{N_t}$ are given in (3).

*Proof:* All the component codes are linear, i.e., for every STBC $\mathcal{C}_n$ each entry of the codeword matrix is a linear combination of the QAM symbols $\{\mathbf{a}(i) | i = 1, \ldots, N_t\}$, and hence for any $\mathbf{X} \in \Delta\mathcal{S}$, there exist $\mathbf{a}_1, \ldots, \mathbf{a}_{N_t} \in \mathbb{Z}[i]^{N_t} \setminus \{\mathbf{0}\}$ and $\mathbf{s}_n = \mathbf{U}\mathbf{a}_n$, $n = 1, \ldots, N_t$ such that

$$\mathbf{X} = \begin{bmatrix} \gamma \mathbf{s}_1(1) & \mathbf{s}_1(2) & \cdots & \mathbf{s}_1(N_t) \\ \mathbf{s}_2(1) & \gamma \mathbf{s}_2(2) & \cdots & \mathbf{s}_2(N_t) \\ \vdots & & \ddots & \vdots \\ \mathbf{s}_{N_t}(1) & \mathbf{s}_{N_t}(2) & \cdots & \gamma \mathbf{s}_{N_t}(N_t) \end{bmatrix},$$

where $\mathbf{s}_n(\ell)$ is the $\ell^{th}$ component of the vector $\mathbf{s}_n$. Since $\mathbf{a}_n \in \mathbb{Z}[i]^{N_t} \setminus \{\mathbf{0}\}$ and $\mathbf{s}_n = \mathbf{U}\mathbf{a}_n$, $\prod_{\ell=1}^{m} |\mathbf{s}(\ell)| > 0$, and hence all the components of $\mathbf{s}_n$ are non-zero. Since $\mathbf{U}$ is an algebraic rotation, and elements of $\mathbb{Z}[i]$ are algebraic, from Lemma 2, all the components of $\mathbf{s}_n$ are algebraic numbers. It follows that all the entries of $\mathbf{X}$ are non-zero, all the off-diagonal entries are algebraic, and all the diagonal entries are products of $e^{\alpha}$ with some algebraic number. Now, the determinant of $\mathbf{X} = [x_{i,j}]$ is $\det(\mathbf{X}) = \sum_{\sigma \in S_{N_t}} \mathrm{sgn}(\sigma) x_{1,\sigma(1)} x_{2,\sigma(2)} \cdots x_{N_t,\sigma(N_t)} =$

$$= \sum_{\sigma \in S_{N_t}} \gamma^{\sum_{n=1}^{N_t} \mathbf{1}(n = \sigma(n))} \mathrm{sgn}(\sigma) \mathbf{s}_1(\sigma(1)) \mathbf{s}_2(\sigma(2)) \cdots \mathbf{s}_{N_t}(\sigma(N_t)),$$
(4)

where $S_{N_t}$ is the set of all permutations on $\{1, \ldots, N_t\}$, $\mathrm{sgn}(\sigma)$ is equal to $1$ or $-1$ if $\sigma$ can be decomposed into even or odd number of transpositions respectively, and $\mathbf{1}(\cdot)$ is the indicator function. From (4) and Lemma 2, $\det(\mathbf{X}) = c_0 + c_1 e^{\alpha} + c_2 e^{2\alpha} + \cdots c_{N_t} e^{N_t \alpha}$, where $c_0, c_1, \ldots, c_{N_t}$ are algebraic. There is exactly one term in (4), corresponding to the identity permutation, that contributes to $\gamma^{N_t}$. Hence, $c_{N_t} = \mathbf{s}_1(1)\mathbf{s}_2(2) \cdots \mathbf{s}_{N_t}(N_t) \neq 0$. Since $0, \alpha, 2\alpha, \ldots, N_t \alpha$ are all distinct and algebraic, and $c_0, \ldots, c_{N_t}$ are algebraic and not all equal to zero, from Theorem 3, we have that $\det(\mathbf{X}) \neq 0$. Thus every $\mathbf{X} \in \Delta\mathcal{S}$ is of full-rank and $\mathsf{r}(\Delta\mathcal{S}) = N_t$, and from Theorem 2, $\mathcal{S}$ achieves full-diversity. ∎

### C. New Finite Feedback Schemes for $T > 1$

*1) Some notations:* The structure of the component codes of the new FFSs for $T > 1$ is similar to the threaded space-time architecture proposed in [24], [25]. Towards describing the new scheme, we first introduce some notations that capture this structure. For any $T > 1$ denote addition modulo $T$ by $\oplus_T$, i.e., for any two integers $a$ and $b$, $a \oplus_T b = (a+b) \mod T$. For a set of $T$ vectors $\mathbf{s}_1, \ldots, \mathbf{s}_T \in \mathbb{C}^{T \times 1}$, we define a $T \times T$ matrix $\mathcal{T}(\mathbf{s}_1,\ldots,\mathbf{s}_T) = [t_{i,j}]$ whose entries are populated by the components of $\mathbf{s}_1,\ldots,\mathbf{s}_T$ as follows. The entries of $\mathcal{T} = [t_{i,j}]$ are partitioned into $T$ *threads*, one corresponding to each of the vectors $\mathbf{s}_1,\ldots,\mathbf{s}_T$. The first thread of $\mathcal{T}$ originates at $t_{1,1}$ and occupies the main diagonal $\{t_{i,i}|i=1,\ldots,T\}$. These entries are populated by the components of the first vector $\mathbf{s}_1$. The second thread originates at $t_{1,2}$ and occupies the entries that are one place to the right of the first thread in $\mathcal{T}$ in cyclic sense. Thus the elements $t_{1,2}, t_{2,3}, \ldots, t_{T-1,T}, t_{T,1}$ form the second thread, and these are populated by the components of the second vector $\mathbf{s}_2$. In general, the $\ell^{th}$ thread originates at $t_{1,\ell}$ and consists of those entries of $\mathcal{T}$ that are one place to the right of the entries of $(\ell-1)^{th}$ thread in cyclic sense. These entries of $\mathcal{T}$ are occupied by the components of the vector $\mathbf{s}_\ell = [\mathbf{s}_\ell(1)\ \mathbf{s}_\ell(2)\ \ldots\ \mathbf{s}_\ell(T)]^T$. Hence, for $1 \leq \ell, i \leq T$ we have

$$t_{i,1+((i-1)\oplus_T(\ell-1))} = \mathbf{s}_\ell(i).$$

*Example 6:* For $T = 3$, we have

$$\mathcal{T}(\mathbf{s}_1,\mathbf{s}_2,\mathbf{s}_3) = [t_{i,j}] = \begin{bmatrix} \mathbf{s}_1(1) & \mathbf{s}_2(1) & \mathbf{s}_3(1) \\ \mathbf{s}_3(2) & \mathbf{s}_1(2) & \mathbf{s}_2(2) \\ \mathbf{s}_2(3) & \mathbf{s}_3(3) & \mathbf{s}_1(3) \end{bmatrix},$$

where the entries occupied by the components of $\mathbf{s}_1$ on the main diagonal form the first thread, the components of $\mathbf{s}_2$ that occupy entries one place to the right of $\mathbf{s}_1$ form the second thread, and the components of $\mathbf{s}_3$ that occupy entries two places to the right of $\mathbf{s}_1$ form the third thread. ■

*Example 7:* The matrix $\mathcal{T}(\mathbf{s}_1,\ldots,\mathbf{s}_4)$, for $\mathbf{s}_1,\ldots,\mathbf{s}_4 \in \mathbb{C}^{4\times 1}$ is

$$\begin{bmatrix} \mathbf{s}_1(1) & \mathbf{s}_2(1) & \mathbf{s}_3(1) & \mathbf{s}_4(1) \\ \mathbf{s}_4(2) & \mathbf{s}_1(2) & \mathbf{s}_2(2) & \mathbf{s}_3(2) \\ \mathbf{s}_3(3) & \mathbf{s}_4(3) & \mathbf{s}_1(3) & \mathbf{s}_2(3) \\ \mathbf{s}_2(4) & \mathbf{s}_3(4) & \mathbf{s}_4(4) & \mathbf{s}_1(4) \end{bmatrix}.$$

■

For any $\mathbf{s} = [\mathbf{s}(1)\ \mathbf{s}(2)\ \cdots\ \mathbf{s}(T)]^T$ and $1 \leq m \leq n \leq T$ we denote the length $n-m+1$ vector $[\mathbf{s}(m)\ \mathbf{s}(m+1)\ \cdots\ \mathbf{s}(n)]^T$ by $\mathbf{s}(m:n)$. If $\mathcal{T}_1,\ldots,\mathcal{T}_N$ are $T \times T$ complex matrices, define

$$\pi\left([\mathcal{T}_1\ \mathcal{T}_2\ \cdots\ \mathcal{T}_{N-1}\ \mathcal{T}_N]\right) = [\mathcal{T}_N\ \mathcal{T}_1\ \mathcal{T}_2\ \cdots\ \mathcal{T}_{N-1}],$$

which is a cyclic shift of the $T \times T$ blocks one place to the right. For any $\mathcal{C} \subset \mathbb{C}^{T \times NT}$, let

$$\pi(\mathcal{C}) = \left\{ \pi\left([\mathcal{T}_1\ \mathcal{T}_2\ \cdots\ \mathcal{T}_N]\right)\ \bigg|\ [\mathcal{T}_1\ \mathcal{T}_2\ \cdots\ \mathcal{T}_N] \in \mathcal{C} \right\}.$$

We now give the construction of new FFSs for $T > 1$.

*2) New FFSs for $T > 1$:* We first give an example of a new FFS for the particular case of $N_t = 4$ antennas with $N = T = 2$. This will help the reader understand the general construction procedure that immediately follows the example.

*Example 8:* Let $\mathcal{A} \subset \mathbb{Z}[i]$ be any QAM constellation, $\mathbf{a}_1, \mathbf{a}_2 \in \mathcal{A}^4$ be vectors of information symbols, and $\mathbf{s}_\ell = \mathbf{U}\mathbf{a}_\ell$, $\ell = 1,2$, where

$$\mathbf{U} = \begin{bmatrix} -0.3664 & -0.7677 & 0.4231 & 0.3121 \\ -0.2264 & -0.4745 & -0.6846 & -0.5050 \\ -0.4745 & 0.2264 & -0.5050 & 0.6846 \\ -0.7677 & 0.3664 & 0.3121 & -0.4231 \end{bmatrix} \quad (5)$$

is a full-diversity algebraic rotation [23]. Let $\beta_1 = i\sqrt{2}, \beta_2 = i\sqrt{3}$ and $\gamma_1 = e^{\beta_1}, \gamma_2 = e^{\beta_2}$. Note that in Example 4 we showed that $\{\beta_1,\beta_2\} = \{i\sqrt{2}, i\sqrt{3}\}$ is linearly independent over $\mathbb{Q}$. The two component STBCs of the proposed FFS are given in (6) and (7) at the top of the next page. Each codeword of $\mathcal{C}_1$ is of the form $[\mathcal{T}_1\ \mathcal{T}_2]$, where

$$\mathcal{T}_1 = \mathcal{T}(\gamma_1 \mathbf{s}_1(1:2), \gamma_2 \mathbf{s}_2(1:2)) \text{ and}$$
$$\mathcal{T}_2 = \mathcal{T}(\mathbf{s}_1(3:4), \mathbf{s}_2(3:4)).$$

The 'threaded' matrix $\mathcal{T}_1$ (respectively $\mathcal{T}_2$) is obtained from the first two entries (last two entries) of $\mathbf{s}_1, \mathbf{s}_2$. Further, the two threads of $\mathcal{T}_1$ are scaled by $\gamma_1$ and $\gamma_2$ respectively. Each codeword of $\mathcal{C}_2$ is of the form $[\mathcal{T}_2\ \mathcal{T}_1] = \pi([\mathcal{T}_1\ \mathcal{T}_2])$. ■

The construction for arbitrary $T$ and $N$ and $N_t = NT$ is as follows. Let $\mathbf{U}$ be an $N_t \times N_t$ full-diversity algebraic rotation, $\mathcal{A} \subset \mathbb{Z}[i]$ be a QAM constellation, $\mathbf{a}_1,\ldots,\mathbf{a}_T \in \mathcal{A}^{N_t}$ be vectors whose components take values independently from $\mathcal{A}$, and $\mathbf{s}_\ell = \mathbf{U}\mathbf{a}_\ell$ for $\ell = 1,\ldots,T$. Further, let $\beta_1,\ldots,\beta_T$ be algebraic numbers that are linearly independent over $\mathbb{Q}$ and $\gamma_\ell = e^{\beta_\ell}$ for $\ell = 1,\ldots,T$. The scalars $\beta_1,\ldots,\beta_T$ can be obtained using Theorem 4 as explained in Section III-A. Now for each $\ell = 1,\ldots,T$, partition the $N_t$-length vector $\mathbf{s}_\ell$ into $N$ vectors $\mathbf{s}_\ell^{(1)}, \mathbf{s}_\ell^{(2)},\ldots,\mathbf{s}_\ell^{(N)}$ of length $T$ each such that

$$\mathbf{s}_\ell = \begin{bmatrix} \mathbf{s}_\ell^{(1)} \\ \mathbf{s}_\ell^{(2)} \\ \vdots \\ \mathbf{s}_\ell^{(N)} \end{bmatrix},$$

i.e., $\mathbf{s}_\ell^{(1)} = \mathbf{s}_\ell(1:T)$, $\mathbf{s}_\ell^{(2)} = \mathbf{s}_\ell(T+1:2T),\ldots$, $\mathbf{s}_\ell^{(N)} = \mathbf{s}_\ell(N_t-T+1:N_t)$. We now construct $N$ matrices $\mathcal{T}_1,\ldots,\mathcal{T}_N$, where $\mathcal{T}_n$ is the threaded $T \times T$ matrix obtained from the $n^{th}$ partitions of $\mathbf{s}_1,\ldots,\mathbf{s}_T$ as follows:

$$\mathcal{T}_1 = \mathcal{T}\left(\gamma_1 \mathbf{s}_1^{(1)}, \gamma_2 \mathbf{s}_2^{(1)},\ldots,\gamma_T \mathbf{s}_T^{(1)}\right), \text{ and}$$
$$\mathcal{T}_n = \mathcal{T}\left(\mathbf{s}_1^{(n)}, \mathbf{s}_2^{(n)},\ldots,\mathbf{s}_T^{(n)}\right), \text{ for } n = 2,\ldots,N.$$

Finally, the $N$ codebooks are

$$\mathcal{C}_1 = \left\{ [\mathcal{T}_1\ \mathcal{T}_2\ \cdots\ \mathcal{T}_N]\ \big|\ \mathbf{a}_1,\ldots,\mathbf{a}_T \in \mathcal{A}^{N_t} \right\}, \text{ and} \quad (8)$$
$$\mathcal{C}_n = \pi(\mathcal{C}_{n-1}),\ n = 2,\ldots,N. \quad (9)$$

*Example 9:* The proposed construction procedure for $T = 2$, $N = 3$ and $N_t = 6$ yields $\mathcal{C}_1, \mathcal{C}_2$ and $\mathcal{C}_3$ as given in (10), (11) and (12) at the top of the next page, where $\mathbf{U}$ is a $6 \times 6$ full-diversity algebraic rotation. ■

If $\beta_1,\ldots,\beta_T$ are purely imaginary, $|\gamma_1| = \cdots = |\gamma_T| = 1$ and for each of the component codes $\mathcal{C}_n$, the average power per each of the transmit antennas is same.

$$\mathcal{C}_1 = \left\{ \begin{bmatrix} \gamma_1 \mathbf{s}_1(1) & \gamma_2 \mathbf{s}_2(1) & \mathbf{s}_1(3) & \mathbf{s}_2(3) \\ \gamma_2 \mathbf{s}_2(2) & \gamma_1 \mathbf{s}_1(2) & \mathbf{s}_2(4) & \mathbf{s}_1(4) \end{bmatrix} \,\bigg|\, \mathbf{s}_1 = \mathbf{U}\mathbf{a}_1, \mathbf{s}_2 = \mathbf{U}\mathbf{a}_2,\ \mathbf{a}_1, \mathbf{a}_2 \in \mathcal{A}^4 \right\} \text{ and} \quad (6)$$

$$\mathcal{C}_2 = \left\{ \begin{bmatrix} \mathbf{s}_1(3) & \mathbf{s}_2(3) & \gamma_1 \mathbf{s}_1(1) & \gamma_2 \mathbf{s}_2(1) \\ \mathbf{s}_2(4) & \mathbf{s}_1(4) & \gamma_2 \mathbf{s}_2(2) & \gamma_1 \mathbf{s}_1(2) \end{bmatrix} \,\bigg|\, \mathbf{s}_1 = \mathbf{U}\mathbf{a}_1, \mathbf{s}_2 = \mathbf{U}\mathbf{a}_2,\ \mathbf{a}_1, \mathbf{a}_2 \in \mathcal{A}^4 \right\}. \quad (7)$$

$$\mathcal{C}_1 = \left\{ \begin{bmatrix} \gamma_1 \mathbf{s}_1(1) & \gamma_2 \mathbf{s}_2(1) & \mathbf{s}_1(3) & \mathbf{s}_2(3) & \mathbf{s}_1(5) & \mathbf{s}_2(5) \\ \gamma_2 \mathbf{s}_2(2) & \gamma_1 \mathbf{s}_1(2) & \mathbf{s}_2(4) & \mathbf{s}_1(4) & \mathbf{s}_2(6) & \mathbf{s}_1(6) \end{bmatrix} \,\bigg|\, \mathbf{s}_1 = \mathbf{U}\mathbf{a}_1, \mathbf{s}_2 = \mathbf{U}\mathbf{a}_2,\ \mathbf{a}_1, \mathbf{a}_2 \in \mathcal{A}^6 \right\}, \quad (10)$$

$$\mathcal{C}_2 = \left\{ \begin{bmatrix} \mathbf{s}_1(5) & \mathbf{s}_2(5) & \gamma_1 \mathbf{s}_1(1) & \gamma_2 \mathbf{s}_2(1) & \mathbf{s}_1(3) & \mathbf{s}_2(3) \\ \mathbf{s}_2(6) & \mathbf{s}_1(6) & \gamma_2 \mathbf{s}_2(2) & \gamma_1 \mathbf{s}_1(2) & \mathbf{s}_2(4) & \mathbf{s}_1(4) \end{bmatrix} \,\bigg|\, \mathbf{s}_1 = \mathbf{U}\mathbf{a}_1, \mathbf{s}_2 = \mathbf{U}\mathbf{a}_2,\ \mathbf{a}_1, \mathbf{a}_2 \in \mathcal{A}^6 \right\}, \quad (11)$$

$$\mathcal{C}_3 = \left\{ \begin{bmatrix} \mathbf{s}_1(3) & \mathbf{s}_2(3) & \mathbf{s}_1(5) & \mathbf{s}_2(5)) & \gamma_1 \mathbf{s}_1(1) & \gamma_2 \mathbf{s}_2(1) \\ \mathbf{s}_2(4) & \mathbf{s}_1(4) & \mathbf{s}_2(6) & \mathbf{s}_1(6) & \gamma_2 \mathbf{s}_2(2) & \gamma_1 \mathbf{s}_1(2) \end{bmatrix} \,\bigg|\, \mathbf{s}_1 = \mathbf{U}\mathbf{a}_1, \mathbf{s}_2 = \mathbf{U}\mathbf{a}_2,\ \mathbf{a}_1, \mathbf{a}_2 \in \mathcal{A}^6 \right\}, \quad (12)$$

*Theorem 5:* If $\mathbf{U}$ is a full-diversity algebraic rotation and $\beta_1, \ldots, \beta_T$ are algebraic numbers that are linearly independent over $\mathbb{Q}$, the FFS $\mathcal{S} = (\mathsf{f}_\mathsf{d}, \mathcal{C}_1, \ldots, \mathcal{C}_N)$ achieves full-diversity, where $\mathcal{C}_1, \ldots, \mathcal{C}_N$ are given by (8) and (9)

*Proof:* See Appendix C for proof. ∎

Since the proposed FFSs encode $K = N_t T$ independent complex symbols they have $R = \frac{K}{T} = N_t$, i.e., full-rate.

For all the new FFSs (both $T = 1$ and $T > 1$), each of the component STBCs is linear, i.e., for each of the STBC $\mathcal{C}_n$, every entry of the codeword matrix is some linear combination of the QAM symbols $\{\mathbf{a}_\ell(i) | \ell = 1, \ldots, T, i = 1, \ldots, N_t\}$. Thus, for a given component code $\mathcal{C}_n$ there exist a set of matrices $\{\mathbf{A}_{\ell,i} | \ell = 1, \ldots, T, i = 1, \ldots, N_t\} \subset \mathbb{C}^{T \times N_t}$ called *linear dispersion* or *weight* matrices [26] such that

$$\mathcal{C}_n = \left\{ \sum_{\ell=1}^{T} \sum_{i=1}^{N_t} \mathbf{a}_\ell(i) \mathbf{A}_{\ell,i} \,\bigg|\, \mathbf{a}_\ell(i) \in \mathcal{A} \right\}.$$

Hence one can use the sphere-decoder [27] to obtain the ML estimate given by (1) [28]. Implementing $\mathsf{f}_\mathsf{d}$, given by (2), requires one to find $\min_{\mathbf{X} \in \Delta \mathcal{C}_n} \|\mathbf{X}\mathbf{H}\|_F^2$ for each $n = 1, \ldots, N$. Again, since $\mathcal{C}_n$ is linear,

$$\Delta \mathcal{C}_n = \left\{ \sum_{\ell=1}^{T} \sum_{i=1}^{N_t} \mathbf{a}_\ell(i) \mathbf{A}_{\ell,i} \,\bigg|\, \mathbf{a}_\ell(i) \in \bar{\Delta}\mathcal{A} \right\} \setminus \{\mathbf{0}\},$$

where $\bar{\Delta}\mathcal{A} = \{a_1 - a_2 | a_1, a_2 \in \mathcal{A}\} \subset \mathbb{Z}[i]$. Hence, finding $\min_{\mathbf{X} \in \Delta \mathcal{C}_n} \|\mathbf{X}\mathbf{H}\|_F^2$ is equivalent to finding the squared norm of the shortest non-zero vector contained in a subset of a lattice. This can be implemented with a minor modification to the sphere-decoding algorithm [29].

## IV. SIMULATION RESULTS

In this section we present simulation results comparing the bit error rate (BER) performance of the new schemes of this paper with the schemes already available in the literature under ML decoding of codewords. In all the simulations, the new FFSs have the best performance while utilizing the least amount of feedback and transmission duration. All the codes discussed in this section use square QAM constellations and Gray encoding to map information bits into QAM symbols.

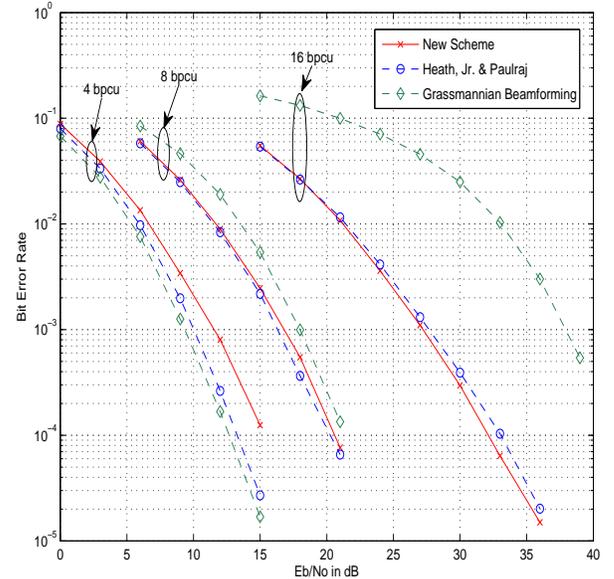

Fig. 1. FFSs for $2 \times 2$ MIMO with $N = 2$.

### A. Schemes for $2 \times 2$ MIMO

In this subsection we compare FFSs for $N_t = N_r = 2$ with $N = 2$-ary feedback. We compare the new FFS of Section II-C1 that was obtained from the Golden code with Grassmannian Beamforming [6] (see Example 1), and the scheme from Heath, Jr. & Paulraj [11]. All three schemes achieve full-diversity, and while the new scheme and Grassmannian Beamforming have $T = 1$ (FT-optimal), the scheme from [11] uses $T = 2$. The new scheme has rate 2 (full-rate), Grassmannian Beamforming has rate 1 and the FFS of [11] uses two codes of different rates: the Alamouti code [14] (rate 1) and spatial multiplexing (rate 2). For bitrate to be constant across the three schemes, if the new FFS uses an $M$-ary QAM constellation, both Grassmannian Beamforming and the Alamouti code for the scheme in [11] use $M^2$-ary

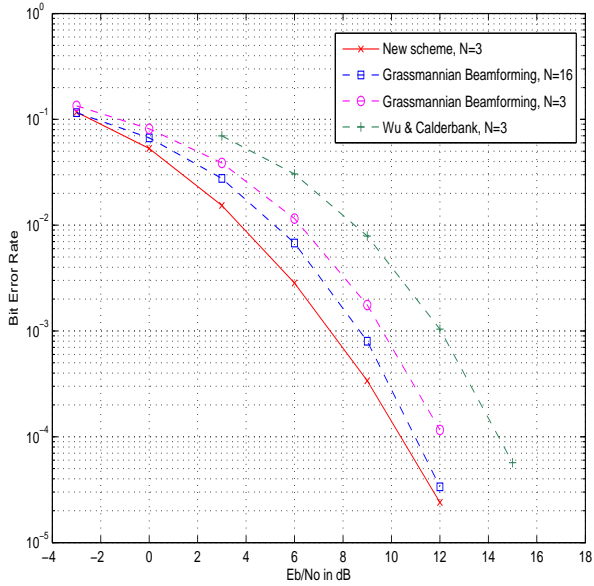

Fig. 2. FFSs for $3 \times 3$ MIMO with 6 bpcu.

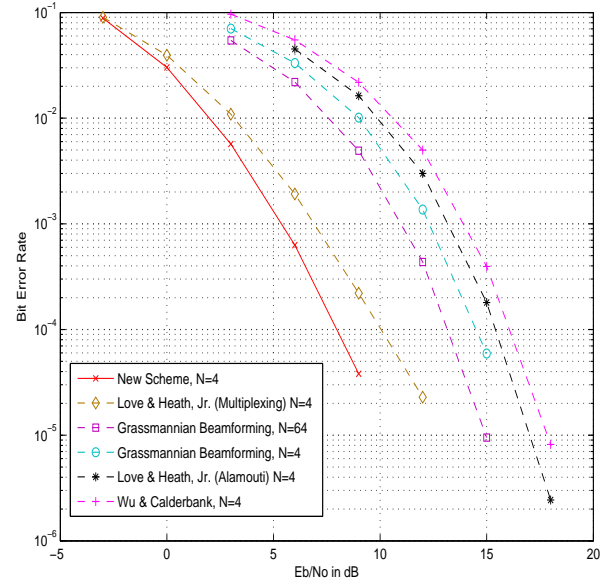

Fig. 3. FFSs for $4 \times 4$ MIMO with 8 bpcu and $N \geq 4$.

QAM, while spatial multiplexing uses $M$-ary QAM. Fig. 1 shows the performance of these three schemes for $4, 8$ and $16$ bpcu. While the new FFS does not fare well for $4$ bpcu, its relative performance improves as the bitrate increases, and for $16$ bpcu it has the lowest BER among the three schemes.

### B. Schemes for $3 \times 3$ MIMO

We now compare the new FFS of Example 5 ($T = 1$ and rate 3) which uses $N = 3$, with Grassmannian Beamforming [6] ($T = 1$ and rate 1) for $N = 3$ and $16$, and the scheme from Wu & Calderbank [12] ($T = N = 3$ and rate 1) for the transmission rate of 6 bpcu. The new code uses $4$-QAM, while the other two schemes use $64$-QAM. The new scheme and the Grassmannian Beamforming that uses $N = 3$ are FT-optimal. Fig. 2 shows the BER performance of the four schemes. We see that the new FFS has the least BER, outperforming even the Grassmannian Beamforming scheme that uses a higher amount of feedback of $N = 16$.

### C. Schemes for $4 \times 4$ MIMO with $N \geq N_t$

We consider the new FFS for $N = 4$, $T = 1$ constructed using the procedure in Section III-B using $\gamma = e^{i\left(\frac{1+\sqrt{5}}{2}\right)}$ and the $4 \times 4$ full-diversity algebraic rotation (5). The new FFS is compared with five other schemes for the bitrate of 8 bpcu: *(i)* the $N = 4, T = 1$ scheme of Love & Heath, Jr. [2] that chooses according the feedback function $f = f_d$ a precoding matrix from a set of $4 \times 2$ matrices to transmit a two-stream spatial multiplexing input over $N_t = 4$ antennas, *(ii)* Grassmannian Beamforming [6] ($T = 1$) with $N = 64$-ary feedback, *(iii)* Grassmannian Beamforming [6] with $N = 4$, *(iv)* the $N = 4, T = 2$ scheme of Love & Heath, Jr. [8] that chooses, based on the feedback index, a precoding matrix from a given set of $4 \times 2$ matrices to transmit an Alamouti code over $N_t = 4$ antennas after precoding, and *(v)* the $N = T = 4$ FFS of Wu & Calderbank [12] for 4 transmit antennas. The new scheme has rate $R = 4$ and uses $4$-QAM constellation. The FFS of [2] has rate $R = 2$ and uses $16$-QAM constellation. The remaining four schemes have rate $R = 1$ and use $256$-QAM. The comparison of BER is shown in Fig. 3, and it is seen that the new FFS has the best performance.

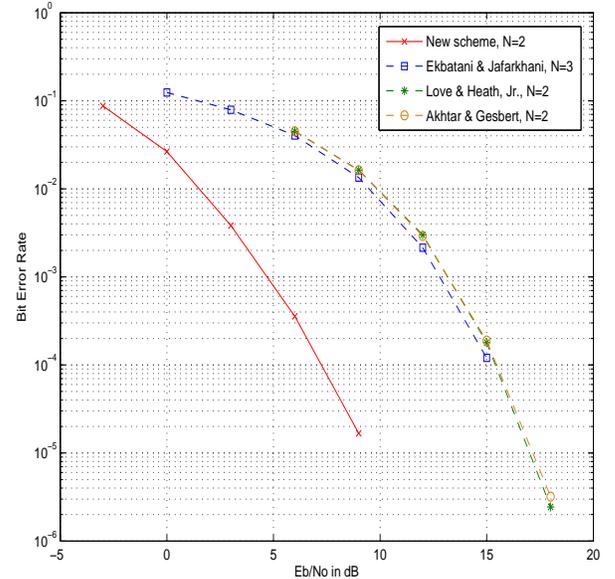

Fig. 4. FFSs for $4 \times 4$ MIMO with 8 bpcu and $N < 4$.

### D. Schemes for $4 \times 4$ MIMO with $N < N_t$

The new scheme considered is the $N = T = 2$ FFS from Example 8. This is compared with: *(i)* the $N = 3, T = 2$

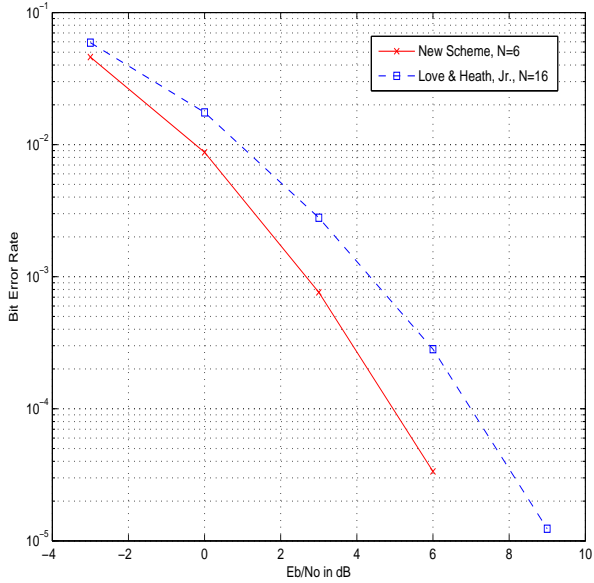

Fig. 5. FFSs for $6 \times 6$ MIMO with 12 bpcu.

scheme of Ekbatani & Jafarkhani [9], *(ii)* the $N = T = 2$ scheme of Love & Heath, Jr. [8], and *(iii)* the $N = T = 2$ scheme of Akhtar & Gesbert [10]. The new scheme has $R = 4$ and uses 4-QAM, while the other three schemes have $R = 1$ and use 256-QAM constellation leading to a bitrate of 8 bpcu. Fig. 4 shows the BER performance of these four schemes.

### E. Schemes for $6 \times 6$ MIMO

We compare the new $N = 6$ FFS obtained from the construction procedure of Section III-B using $\gamma = e^{i\left(\frac{1+\sqrt{5}}{2}\right)}$ and the $6 \times 6$ full-diversity algebraic rotation labeled 'mixed 2x3' in [23]. This is compared with the rate 3 FFS of [8] that uses $f = f_d$ and $N = 16$-ary feedback. The new FFS uses 4-QAM while the scheme from [8] uses 16-QAM, both leading to 12 bpcu. Fig. 5 shows the BER performance of these two schemes, and we see that while using less amount of feedback the new scheme outperforms the scheme from [8].

## V. CONCLUSION

In this paper we have given a universal necessary condition for any FFS to achieve full-diversity in a Rayleigh block fading channel with finite noise-free delay-free feedback. Based on this criterion we have introduced the notion of FT-optimal schemes that use minimum feedback for the given transmission duration and minimum transmission duration for the given feedback to achieve full-diversity. We have also given a sufficient condition for full-diversity for those schemes in which the receiver chooses the component STBC whose minimum Euclidean distance is maximum. Based on this criterion and using tools from algebraic number theory, we have constructed full-rate full-diversity FT-optimal FFSs for all triples $(N, T, N_t)$ with $N_t = NT$. These are the first full-rate full-diversity FFSs reported in the literature for $T < N_t$. Through simulation results we showed that the proposed FFSs have the best performance among the schemes available in the literature. Following are some of the questions that are yet to be addressed.

- Though the necessary condition presented in Section II-B for full-diversity is universal, the sufficient condition of Section II-C applies to only those FFSs that use $f = f_d$. Is there a universal necessary and sufficient criterion for full-diversity?
- Finding $f_d(\mathbf{H})$ at the receiver is equivalent to solving the closest lattice point problem for $N$ different lattices, and hence this operation is of high complexity. Are there feedback functions that can be implemented with low complexity and still lead to full-diversity? Can one design the component STBCs in such a way that $f_d$ itself can be implemented with low complexity?

## APPENDIX A
## PROOF OF THEOREM 1

Let $\mathbf{X} \in \Delta \mathcal{S}$ be of rank $\mathsf{r}(\Delta \mathcal{S})$. There exist $\mathbf{X}_a(n), \mathbf{X}_b(n) \in \mathcal{C}_n$, $n = 1, \ldots, N$, such that

$$\mathbf{X} = \begin{bmatrix} \mathbf{X}_a(1) - \mathbf{X}_b(1) \\ \mathbf{X}_a(2) - \mathbf{X}_b(2) \\ \vdots \\ \mathbf{X}_a(N) - \mathbf{X}_b(N) \end{bmatrix}.$$

Let the codebook size $|\mathcal{C}_1| = \cdots = |\mathcal{C}_N| = M$. For a fixed channel realization $\mathbf{H}$, if the feedback index $\mathsf{f}(\mathbf{H}) = n$, then the probability of codeword error of the ML decoder when $\mathbf{X}_a(n)$ is transmitted is lower bounded by the pairwise error probability $\mathsf{PEP}(\mathbf{X}_a(n) \to \mathbf{X}_b(n)|\mathbf{H})$ between the codewords $\mathbf{X}_a(n), \mathbf{X}_b(n)$. Hence we have $\mathsf{P}_e(\mathbf{H})$

$$\geq \mathsf{P}(\mathbf{X}_a(n) \text{ is transmitted}|\mathbf{H})\mathsf{PEP}(\mathbf{X}_a(n) \to \mathbf{X}_b(n)|\mathbf{H})$$

$$= \frac{1}{M} Q\left(\sqrt{\frac{E}{2N_0}} \|(\mathbf{X_a}(n) - \mathbf{X}_b(n))\mathbf{H}\|_F\right),$$

where $Q(\cdot)$ is the Gaussian tail function. Since $\|(\mathbf{X_a}(n) - \mathbf{X}_b(n))\mathbf{H}\|_F \leq \|\mathbf{X}\mathbf{H}\|_F$ and $Q$ is a monotonically decreasing function, we have

$$\mathsf{P}_e(\mathbf{H}) \geq \frac{1}{M} Q\left(\sqrt{\frac{E}{2N_0}} \|\mathbf{X}\mathbf{H}\|_F\right). \tag{13}$$

From [30], for any $\beta > 1$ and $0 < \alpha < \sqrt{\frac{2e}{\pi}} \frac{\sqrt{\beta-1}}{\beta}$, we have $Q(x) \geq \frac{\alpha}{2}\exp(-\frac{\beta x^2}{2})$. Using $\alpha = \frac{1}{2}$ and $\beta = 2$ to lower bound the right hand side of (13), we get

$$\mathsf{P}_e(\mathbf{H}) \geq \frac{1}{4M}\exp\left(-\frac{E}{2N_0}\|\mathbf{X}\mathbf{H}\|_F^2\right). \tag{14}$$

Now, $\|\mathbf{X}\mathbf{H}\|_F^2 = \mathrm{tr}(\mathbf{H}^H \mathbf{X}^H \mathbf{X} \mathbf{H})$. Let $\mathbf{X}^H \mathbf{X} = \mathbf{U}\mathbf{D}\mathbf{U}^H$ be the eigen decomposition of $\mathbf{X}^H \mathbf{X}$, where $\mathbf{U} \in \mathbb{C}^{N_t \times N_t}$ is unitary and $\mathbf{D}$ is the diagonal matrix consisting of the eigenvalues of $\mathbf{X}^H \mathbf{X}$. Let $\lambda_1, \lambda_2, \ldots, \lambda_{\mathsf{r}(\Delta \mathcal{S})}$ be the non-zero eigenvalues of $\mathbf{X}^H \mathbf{X}$ and $\widetilde{\mathbf{H}} = \mathbf{U}^H \mathbf{H}$, then $\|\mathbf{X}\mathbf{H}\|_F^2 =$

$\text{tr}(\widetilde{\mathbf{H}}^H \mathbf{D} \widetilde{\mathbf{H}}) = \sum_{j=1}^{N_r} \sum_{i=1}^{r(\Delta \mathcal{S})} \lambda_i |\tilde{h}_{i,j}|^2$, where $\widetilde{\mathbf{H}} = [\tilde{h}_{i,j}]$. Since $\widetilde{\mathbf{H}}$ and $\mathbf{H}$ are identically distributed, the variables $|\tilde{h}_{i,j}|^2$ are independent and identically distributed exponential random variables with unit mean. Averaging (14) with respect to $\mathbf{H}$ we get $\mathsf{P}_e =$

$$\mathsf{E}(\mathsf{P}_e(\mathbf{H})) \geq \frac{1}{4M} \mathsf{E}\left(\exp\left(-\frac{E}{2N_0} \sum_{j=1}^{N_r} \sum_{i=1}^{r(\Delta \mathcal{S})} \lambda_i |\tilde{h}_{i,j}|^2\right)\right)$$

$$= \frac{1}{4M} \prod_{j=1}^{N_r} \prod_{i=1}^{r(\Delta \mathcal{S})} \mathsf{E}\left(\exp\left(-\frac{E}{2N_0} \lambda_i |\tilde{h}_{i,j}|^2\right)\right)$$

$$= \frac{1}{4M} \prod_{j=1}^{N_r} \prod_{i=1}^{r(\Delta \mathcal{S})} \left(1 + \frac{\lambda_i E}{2N_0}\right)^{-1}.$$

The last equality is due to the fact that for an exponentially distributed random variable $x$ with unit mean, and for any $s > 0$, $\mathsf{E}(\exp(-sx)) = (1+s)^{-1}$. For large values of $\frac{E}{N_0}$, we have

$$\mathsf{P}_e \gtrsim \frac{1}{4M} \left(\frac{E}{2N_0}\right)^{-r(\Delta \mathcal{S}) N_r} \prod_{i=1}^{r(\Delta \mathcal{S})} \lambda_i^{N_r}.$$

Hence the probability of error decays at the most as fast as $\left(\frac{E}{N_0}\right)^{-r(\Delta \mathcal{S}) N_r}$. This completes the proof.

## APPENDIX B
## PROOF OF THEOREM 2

Let $|\mathcal{C}_1| = \cdots = |\mathcal{C}_N| = M$, and let the codewords of each codebook $\mathcal{C}_n$ be indexed by the message index $m \in \{1, \ldots, M\}$, i.e., let $\mathcal{C}_n = \{\mathbf{X}_m(n) | m \in \{1, \ldots, M\}\}$. In order to prove the theorem, we derive an upper bound on the pairwise error probability $\mathsf{PEP}(m_1 \to m_2)$ between any two distinct message indices $m_1, m_2 \in \{1, \ldots, M\}$. For a given channel realization $\mathbf{H}$, let $\mathsf{f}_\mathsf{d}(\mathbf{H}) = n^*$, then

$$\mathsf{PEP}(m_1 \to m_2 | \mathbf{H})$$
$$= Q\left(\sqrt{\frac{E}{2N_0}} \|(\mathbf{X}_{m_1}(n^*) - \mathbf{X}_{m_2}(n^*)) \mathbf{H}\|_F\right).$$

Using the Chernoff bound [30] $Q(x) \leq \frac{1}{2} \exp(-\frac{x^2}{2})$, we get

$$\mathsf{PEP}(m_1 \to m_2 | \mathbf{H})$$
$$\leq \frac{1}{2} \exp\left(-\frac{E}{4N_0} \|(\mathbf{X}_{m_1}(n^*) - \mathbf{X}_{m_2}(n^*)) \mathbf{H}\|_F^2\right). \quad (15)$$

For each $n = 1, \ldots, N$, let $\mathbf{X}_{min}(n) = \arg\min_{\mathbf{X} \in \Delta \mathcal{C}_n} \|\mathbf{X}\mathbf{H}\|_F^2$, and

$$\mathbf{X}_{min} = \begin{bmatrix} \mathbf{X}_{min}(1) \\ \mathbf{X}_{min}(2) \\ \vdots \\ \mathbf{X}_{min}(N) \end{bmatrix}.$$

Note that $\|\mathbf{X}_{min}\mathbf{H}\|_F^2 \geq \lambda_{N_t}(\mathbf{X}_{min}^H \mathbf{X}_{min}) \|\mathbf{H}\|_F^2$, where $\lambda_{N_t}(\mathbf{X}_{min}^H \mathbf{X}_{min})$ is the smallest singular value of $\mathbf{X}_{min}^H \mathbf{X}_{min}$. Let $\lambda^* = \min_{\mathbf{X} \in \Delta \mathcal{S}} \lambda_{N_t}(\mathbf{X}^H \mathbf{X})$. Since all the matrices in $\Delta \mathcal{S}$ have rank $N_t$, we have $\lambda^* > 0$, and

$$\|\mathbf{X}_{min}\mathbf{H}\|_F^2 \geq \lambda_{N_t}(\mathbf{X}_{min}^H \mathbf{X}_{min}) \|\mathbf{H}\|_F^2 \geq \lambda^* \|\mathbf{H}\|_F^2. \quad (16)$$

Since $n^* = \arg\max_{n \in \{1, \ldots, N\}} \|\mathbf{X}_{min}(n)\mathbf{H}\|_F^2$, we have

$$\|\mathbf{X}_{min}(n^*)\mathbf{H}\|_F^2 \geq \frac{1}{N} \sum_{n=1}^{N} \|\mathbf{X}_{min}(n)\mathbf{H}\|_F^2 = \frac{1}{N} \|\mathbf{X}_{min}\mathbf{H}\|_F^2. \quad (17)$$

From (16) and (17) we have

$$\|(\mathbf{X}_{m_1}(n^*) - \mathbf{X}_{m_2}(n^*))\mathbf{H}\|_F^2 \geq \|\mathbf{X}_{min}(n^*)\mathbf{H}\|_F^2$$
$$\geq \frac{1}{N} \|\mathbf{X}_{min}\mathbf{H}\|_F^2$$
$$\geq \frac{\lambda^*}{N} \|\mathbf{H}\|_F^2.$$

Thus, we can upper bound the left hand side of (15) as

$$\mathsf{PEP}(m_1 \to m_2 | \mathbf{H}) \leq \frac{1}{2} \exp\left(-\frac{E\lambda^*}{4NN_0} \|\mathbf{H}\|_F^2\right)$$
$$= \frac{1}{2} \prod_{i=1}^{N_t} \prod_{j=1}^{N_r} \exp\left(-\frac{E\lambda^*}{4NN_0} |h_{i,j}|^2\right), \quad (18)$$

where $\mathbf{H} = [h_{i,j}]$, and the variables $|h_{i,j}|^2$ are independent random variables that are exponentially distributed with unit mean. Averaging (18) with respect to $\mathbf{H}$, we obtain

$$\mathsf{PEP}(m_1 \to m_2) \leq \frac{1}{2} \prod_{i=1}^{N_t} \prod_{j=1}^{N_r} \mathsf{E}\left(\exp\left(-\frac{E\lambda^*}{4NN_0} |h_{i,j}|^2\right)\right)$$
$$= \frac{1}{2} \left(1 + \frac{E\lambda^*}{4NN_0}\right)^{-N_t N_r}$$

For large values of $\frac{E}{N_0}$ we have

$$\mathsf{PEP}(m_1 \to m_2) \lesssim \frac{1}{2} \left(\frac{E\lambda^*}{4NN_0}\right)^{-N_t N_r}.$$

This completes the proof.

## APPENDIX C
## PROOF OF THEOREM 5

Let $\mathbf{X} = [\mathbf{X}_1^T \ \mathbf{X}_2^T \ \cdots \ \mathbf{X}_N^T]^T \in \Delta \mathcal{S}$. Since the codes $\mathcal{C}_1, \ldots, \mathcal{C}_N$ are linear, for each $n \in \{1, \ldots, N\}$ there exist vectors $\mathbf{a}_1, \ldots, \mathbf{a}_T \in \mathbb{Z}[i]^{N_t}$, not all zero, such that

$$\mathbf{X}_n = \pi^{(n-1)}([\mathcal{T}_1 \ \mathcal{T}_2 \ \cdots \ \mathcal{T}_N]), \text{ where}$$

$\mathcal{T}_1 = \mathcal{T}(\gamma_1 \mathbf{s}_1^{(1)}, \ldots, \gamma_T \mathbf{s}_T^{(1)})$ and $\mathcal{T}_m = \mathcal{T}(\mathbf{s}_1^{(m)}, \ldots, \mathbf{s}_T^{(m)})$ for $m > 1$. All the entries of $\mathcal{T}_m$, $m > 1$, are algebraic, and each entry of $\mathcal{T}_1$ is either 0 or a product $\gamma_\ell \alpha$ for some $\ell \in \{1, \ldots, T\}$ and some algebraic number $\alpha$. Hence the determinant of $\mathbf{X}$ is a polynomial $f(x_1, \ldots, x_T)$ with algebraic coefficients and degree at the most $N_t$ with respect to each $x_\ell$, evaluated at the point $(x_1, \ldots, x_T) = (\gamma_1, \ldots, \gamma_T)$. Let $\mathbb{Z}_{N_t+1} = \{0, 1, \ldots, N_t\}$, and for any $\mathbf{p} \in \mathbb{Z}_{N_t+1}^T$ let $\gamma^{\mathbf{p}}$ denote the product $\gamma_1^{\mathbf{p}(1)} \gamma_2^{\mathbf{p}(2)} \cdots \gamma_T^{\mathbf{p}(T)}$. Then $\det(\mathbf{X}) = \sum_{\mathbf{p} \in \mathbb{Z}_{N_t+1}^T} c_{\mathbf{p}} \gamma^{\mathbf{p}}$, where the scalars $c_{\mathbf{p}}$ are algebraic. In order to use Theorem 3 we need to show that all the $\gamma_{\mathbf{p}}$'s are distinct and at least one of the $c_{\mathbf{p}}$ is non-zero. Suppose $\mathbf{p}_1, \mathbf{p}_2 \in \mathbb{Z}_{N_t+1}^T$ are distinct. We have

$$\gamma^{\mathbf{p}_1} = e^{\sum_{\ell=1}^T \beta_\ell \mathbf{p}_1(\ell)} \text{ and } \gamma^{\mathbf{p}_2} = e^{\sum_{\ell=1}^T \beta_\ell \mathbf{p}_2(\ell)}.$$

Since $\mathbf{p}_1, \mathbf{p}_2 \in \mathbb{Q}^{T\times 1}$ are distinct, and $\{\beta_1, \ldots, \beta_T\}$ is linearly independent over $\mathbb{Q}$ we have $\sum_{\ell=1}^T \beta_\ell \mathbf{p}_1(\ell) \neq \sum_{\ell=1}^T \beta_\ell \mathbf{p}_2(\ell)$. Thus $\gamma^{\mathbf{p}_1}$ and $\gamma^{\mathbf{p}_2}$ are distinct for all pairs of distinct $\mathbf{p}_1, \mathbf{p}_2$. Now, using Theorems 2 and 3, it is enough to show that $c_{\mathbf{p}} \neq 0$ for some $\mathbf{p} \in \mathbb{Z}_{N_t+1}^T$.

Partition the matrix $\mathbf{X}$ into $T \times T$ matrices $\mathbf{X}^{(i,j)}$ such that

$$\mathbf{X} = \begin{bmatrix} \mathbf{X}^{(1,1)} & \mathbf{X}^{(1,2)} & \cdots & \mathbf{X}^{(1,N)} \\ \mathbf{X}^{(2,1)} & \mathbf{X}^{(2,2)} & \cdots & \mathbf{X}^{(2,N)} \\ \vdots & & \ddots & \vdots \\ \mathbf{X}^{(N,1)} & \mathbf{X}^{(N,2)} & \cdots & \mathbf{X}^{(N,N)} \end{bmatrix}.$$

For $i \neq j$, every entry of $\mathbf{X}^{(i,j)}$ is algebraic. Since $\mathbf{U}$ is a full-diversity rotation, for every $i \in \{1, \ldots, N\}$ and $\ell \in \{1, \ldots, T\}$, either all the entries of the $\ell^{th}$ thread of $\mathbf{X}^{(i,i)}$ are zero or every entry of the $\ell^{th}$ thread is non-zero. In the latter case each such entry is a product of $\gamma_\ell$ with some algebraic number. From among $\mathbf{X}^{(1,1)}, \mathbf{X}^{(2,2)}, \ldots, \mathbf{X}^{(N,N)}$, let $m_1$ be the number of matrices whose first thread is non-zero. Let $m_2$ be the number of matrices whose first thread is zero and second thread is non-zero. And in general, let $m_\ell$ be the number of matrices whose first $\ell - 1$ threads are zero and the $\ell^{th}$ thread is non-zero. Since for each $\mathbf{X}^{(i,i)}$ at least one of the $T$ threads is non-zero, we have $m_1 + \cdots + m_T = N$, and $m_1 T + m_2 T + \cdots + m_T T = N_t$. To complete the proof, we will now show that for $\mathbf{p}^* = [m_1 T \ m_2 T \ \cdots \ m_T T]^T$, we have $c_{\mathbf{p}^*} \neq 0$.

Writing $\mathbf{X} = [x_{s,t}]$, we have

$$\sum_{\mathbf{p} \in \mathbb{Z}_{N_t+1}^T} c_{\mathbf{p}} \gamma^{\mathbf{p}} = \det(\mathbf{X})$$
$$= \sum_{\sigma \in S_{N_t}} \mathsf{sgn}(\sigma) x_{1,\sigma(1)} x_{2,\sigma(2)} \cdots x_{N_t, \sigma(N_t)}, \quad (19)$$

where $S_{N_t}$ is the set of all permutations on $\{1, \ldots, N_t\}$. Each term in summation in (19) is of the form $\alpha \gamma^{\mathbf{p}}$, where $\alpha$ is algebraic and $\mathbf{p} \in \mathbb{Z}_{N_t+1}^T$. Let $\sigma \in S_{N_t}$ be any permutation associated with $\mathbf{p}^*$ that contributes a non-zero term to (19), i.e., $\mathsf{sgn}(\sigma) x_{1,\sigma(1)} x_{2,\sigma(2)} \cdots x_{N_t, \sigma(N_t)} =$

$$\alpha \gamma^{\mathbf{p}^*} = \alpha \gamma_1^{m_1 T} \gamma_2^{m_2 T} \cdots \gamma_T^{m_T T}.$$

Since every $m_1 T + \cdots + m_T T = N_t$, for every $s \in \{1, \ldots, N_t\}$, $x_{s,\sigma(s)}$ is a product of an algebraic number and one of the $\gamma_\ell$'s, i.e., each $x_{s,\sigma(s)}$ is an entry of one the matrices $\mathbf{X}^{(1,1)}, \ldots, \mathbf{X}^{(N,N)}$. Hence, there exist $N$ permutations: $\sigma_1$ on $\{1, \ldots, T\}$, $\sigma_2$ on $\{T+1, \ldots, 2T\}$, $\ldots$, and $\sigma_N$ on $\{N_t - T + 1, \ldots, N_t\}$ such that

$$\alpha \gamma_1^{m_1 T} \gamma_2^{m_2 T} \cdots \gamma_T^{m_T T} = \mathsf{sgn}(\sigma) x_{1,\sigma(1)} x_{2,\sigma(2)} \cdots x_{N_t, \sigma(N_t)}$$
$$= \prod_{n=1}^N \prod_{i=(n-1)T+1}^{nT} x_{i,\sigma_n(i)}. \quad (20)$$

For $\ell = 1, \ldots, T$, let $\mathcal{I}_\ell \subseteq \{1, \ldots, N\}$ be set of the indices of those matrices in $\mathbf{X}^{(1,1)}, \ldots, \mathbf{X}^{(N,N)}$ whose first $(\ell - 1)$ threads are zero, and the $\ell^{th}$ thread is non-zero. Since the degree of $\gamma_1$ in (20) is $m_1 T$ and since there are only $m_1 T$ non-zero entries in $\mathbf{X}$ that contain terms of type $\zeta \gamma_1$, $\zeta$ algebraic, and all of them are contained in the diagonal blocks indexed by elements in $\mathcal{I}_1$, it follows that for every $n \in \mathcal{I}_1$, $\sigma_n$ is the identity map on $\{(n-1)T+1, \ldots, nT\}$. There are only $m_2 T$ non-zero entries in $\mathbf{X}$, outside the blocks indexed by elements of $\mathcal{I}_1$, of the type $\zeta \gamma_2$, $\zeta$ algebraic, and these are contained in the block matrices whose indices belong to $\mathcal{I}_2$. Since the degree of $\gamma_2$ is $m_2 T$ in (20), for every $n \in \mathcal{I}_2$, $\sigma_n(i) = (n-1)T + 1 + ((i - (n-1)T - 1) \oplus_T 1)$. Extending this argument, for any $\ell > 1$, there are only $m_\ell T$ non-zero entries in $\mathbf{X}$ that are of the form $\zeta \gamma_\ell$, outside of the blocks $\mathbf{X}^{(i,i)}$, $i \in \mathcal{I}_1 \cup \cdots \cup \mathcal{I}_{\ell-1}$, and these are contained in the matrices $\mathbf{X}^{(i,i)}$, $i \in \mathcal{I}_\ell$. Since the degree of $\gamma_\ell$ in (20) is $m_\ell T$, for every $n \in \mathcal{I}_\ell$ we have

$$\sigma_n(i) = (n-1)T + 1 + ((i - (n-1)T - 1) \oplus_T (\ell - 1)),$$

for $i \in \{(n-1)T + 1, \ldots, nT\}$. Thus, there exists a unique $\sigma \in S_{N_t}$ that contributes a non-zero term of type $\alpha \gamma^{\mathbf{p}^*}$, $\alpha$ algebraic, to the sum (19). Hence $c_{\mathbf{p}^*} \neq 0$, and this completes the proof.